\documentclass[preprint]{emulateapj}
\usepackage{natbib}
\usepackage{graphicx}

\begin{document} 

\title{Depletion of molecular gas by an accretion outburst in a protoplanetary disk} 

\author{A. Banzatti\altaffilmark{1}, K. M. Pontoppidan\altaffilmark{1}, S. Bruderer\altaffilmark{2}, J. Muzerolle\altaffilmark{1},  M. R. Meyer\altaffilmark{3}} 
\altaffiltext{1}{Space Telescope Science Institute, Baltimore, MD 21218, USA} 
\altaffiltext{2}{Max-Planck-Institut f\"ur Extraterrestrische Physik, Giessenbachstr. 1, D-85748 Garching bei M\"unchen, Germany} 
\altaffiltext{3}{ETH Z\"urich, Institut f\"ur Astronomie, Wolfgang-Pauli-Strasse 27, CH-8093 Z\"urich, Switzerland} 

\textit{Accepted by the Astrophysical Journal Letters}

\email{banzatti@stsci.edu}

\begin{abstract} 
We investigate new and archival 3--5\,$\mu$m high resolution ($\sim3$\,km\,s$^{-1}$) spectroscopy of molecular gas in the inner disk of the young solar-mass star EX Lupi, taken during and after the strong accretion outburst of 2008. The data were obtained using the CRIRES spectrometer at the ESO Very Large Telescope in 2008 and 2014. In 2008, emission lines from CO, H$_{2}$O, and OH were detected with broad profiles tracing gas near and within the corotation radius (0.02--0.3\,AU). In 2014, the spectra display marked differences. The CO lines, while still detected, are much weaker, and the H$_{2}$O and OH lines have disappeared altogether. At 3\,$\mu$m a veiled stellar photospheric spectrum is observed. Our analysis finds that the molecular gas mass in the inner disk has decreased by an order of magnitude since the outburst, matching a similar decrease in the accretion rate onto the star.
We discuss these findings in the context of a rapid depletion of material accumulated beyond the disk corotation radius during quiescent periods, as proposed by models of episodic accretion in EXor type young stars.
\end{abstract}


\section{INTRODUCTION} \label{sec:intr}
\begin{figure*}
\centering
\includegraphics[width=1\textwidth]{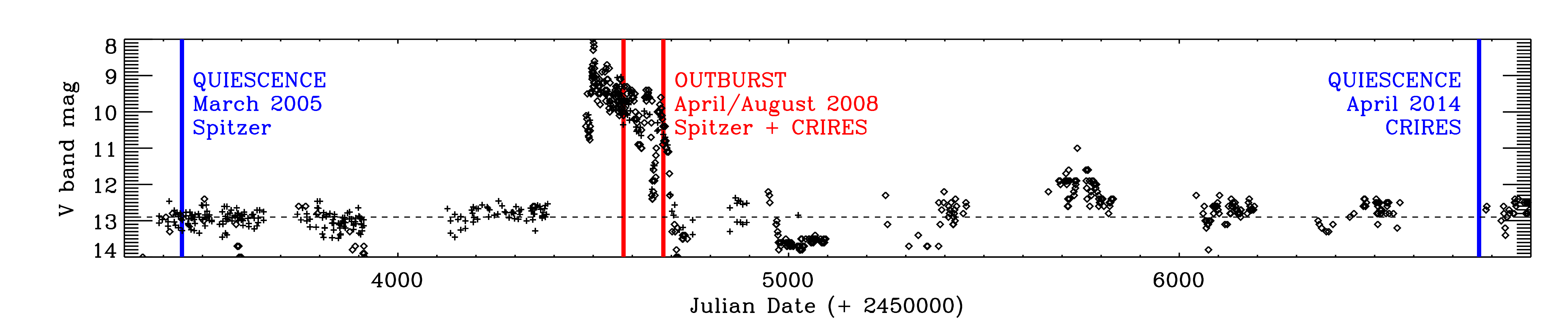} 
\includegraphics[width=1\textwidth]{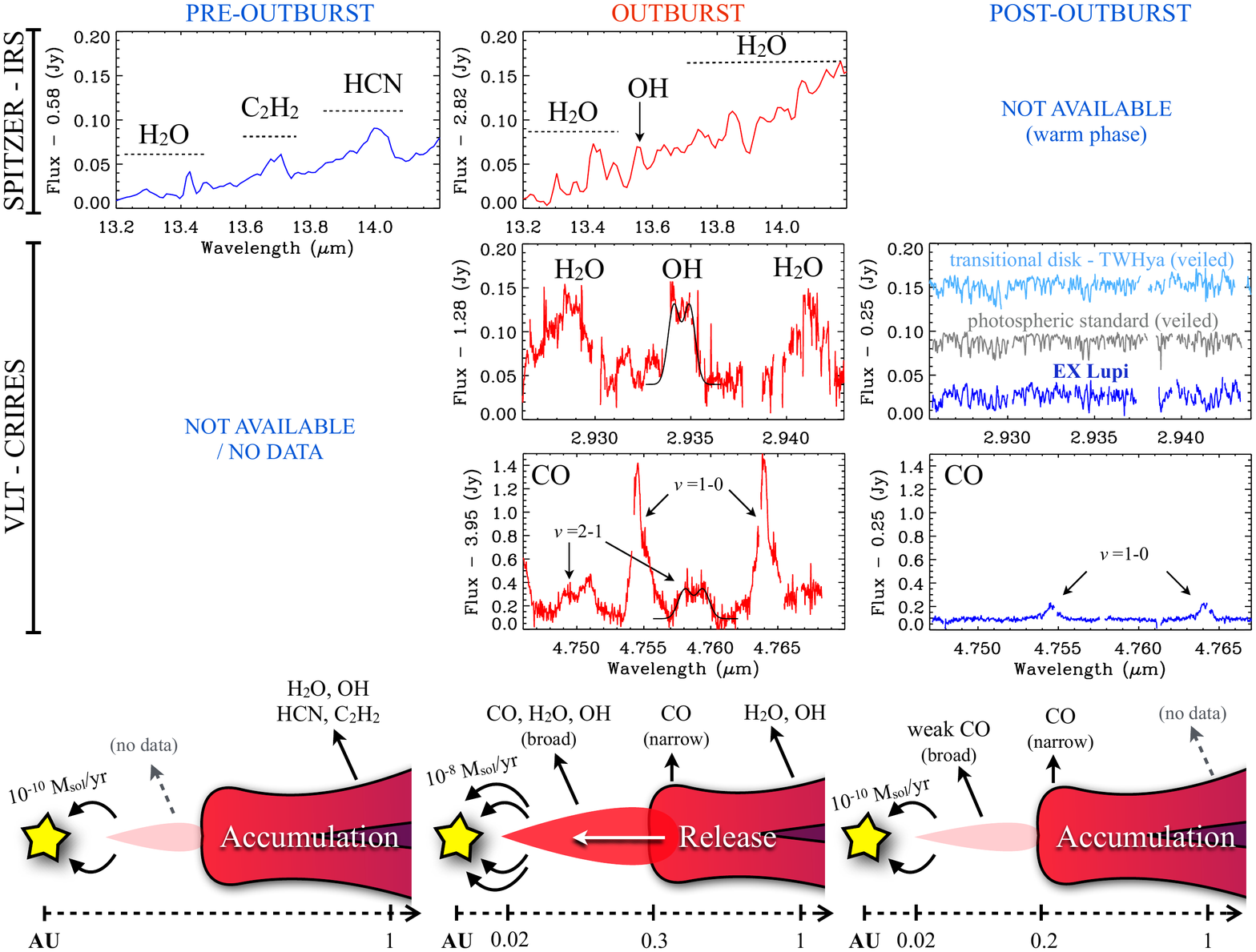} 
\caption{Overview of infrared molecular line emission changes monitored in EX Lupi before, during, and after the 2008 outburst. \textit{Top}: the V-band lightcurve of EX Lupi, combining data from the ASAS database \citep[crosses,][]{asas} and data from the AAVSO database (diamonds, www.aavso.org). \textit{Middle}: portions of EX Lupi spectra from \textit{Spitzer}-IRS \citep{banz12} and VLT-CRIRES (this work). \textit{Bottom}: cartoon illustrating the inner disk in EX Lupi during the three phases of accretion, highlighting the molecular gas location.}
\label{fig:proposal}
\end{figure*}

During star formation, the rate of accretion of circumstellar disk material onto the pre-main sequence star is not constant. Sharp order-of-magnitude enhancements in the mass accretion rate called \textit{accretion outbursts} have long been known to occur in young stars \citep{herb77}. Historically, two classes of eruptive young stars have been observationally identified. FUor (FUOri-type) systems are characterized by bursts in accretion rate up to $\sim10^{-5}$\,M$_{\odot}$\,yr$^{-1}$ and duration of 100s years, while EXLupi-type systems \citep[called ``EXors'' by][]{herb89} show episodic outbursts that are a factor of $\approx100$ weaker both in accretion rate and in duration. These accretion outbursts are thought to be linked to the evolution of protoplanetary disks. In fact, the inner region of the disk directly feeds the central star, and both its energy budget and its chemistry are modified by high-energy radiation such as that generated by accretion shocks \citep[e.g.][]{hartm,audard14}. While FUors have longer been known, EXors have recently attracted increasing attention due to their shorter accretion cycle timescale that provides exceptional opportunities for observing recurring accretion events and ongoing physical and chemical processing of disk material \citep[][and Figure \ref{fig:proposal}]{abra09,banz12}. 

A theoretical picture of episodic accretion in EXors is emerging in which mass accumulates outside the star-disk corotation radius until the surface density increases enough to overcome the centrifugal barrier, perturbing the magnetospheric radius inward of corotation and bringing the inner disk into a high-accretion phase until the built-up material is depleted \citep{spruit,dAng10,dAng11,dAng12}. Determining the amount and extent of gas in the inner disk before, during, and after an outburst can therefore provide unique constraints to episodic accretion models and help understand the basic physics governing the evolution of inner regions in accreting protoplanetary disks.

In this paper, we compare new and archival near-infrared (NIR) spectroscopy of molecular gas (CO, H$_{2}$O, and OH) during and after the 2008 outburst of EX Lupi.
EX Lupi is a 1--3\,Myr old T Tauri star with $M_*=0.6\,M_{\odot}$, $R_*=1.6\,R_{\odot}$, and spectral type M0 \citep{grasvel05}, at a distance of 155 pc in the Lupus complex \citep{lomb}. It is the prototypical EXor variable \citep{herb89,herb01}, where the record of episodic accretion flares dates back to 1901 \citep{McL}. In 2008 the strongest outburst on record occurred with a 5 mag increase in the V band (Figure \ref{fig:proposal}). On that occasion, EX Lupi was observed over a broad wavelength range from the X-ray and UV \citep{grosso}, to the optical \citep{asp,sicagu12} and the infrared \citep{goto,kosp11} using a plethora of telescopes. With this work, we build upon the heritage of NIR observations taken in 2008. Using new NIR spectroscopy obtained in 2014, we find that the inner disk is depleted by one order of magnitude in molecular gas mass, unveiling a forest of stellar photosphere lines at 3\,$\mu$m.

\section{CRIRES SPECTROSCOPY OF EXLupi AT 3 AND 5 $\mu$m} \label{sec:data}
We present velocity-resolved infrared spectroscopy of EX Lupi using the CRyogenic Infrared Echelle Spectrometer \citep[CRIRES,][]{crires} on the Very Large Telescope of the European Southern Observatory. New data were taken in April 2014, covering the spectral regions 2.91--3.07\,$\mu$m and 4.66--4.89\,$\mu$m (program 093.C-0432). Similar spectral regions had been observed during the accretion outburst in 2008, in April (2.91--2.99\,$\mu$m) and in August (4.66--4.90 $\mu$m), as part of the ESO Large Program 179.C-0151 \citep[][]{pont_msgr}. In this work, we compare spectral regions common to the two epochs: 2.91--2.96\,$\mu$m (henceforth referred to as the 3\,$\mu$m setting) and 4.66--4.89\,$\mu$m (henceforth referred to as the 5\,$\mu$m setting). The two epochs were taken using identical instrumental parameters. The slit width was $\sim0.2"$, providing spectral resolution of $\sim3.2$\,km\,s$^{-1}$, and the on-source integration time was 20 min per spectral setting. Early-type telluric standard stars were observed at similar airmasses to EX Lupi (typically within 0.05, and always less than 0.1 difference): HR5812 at 3\,$\mu$m, BS6175 and HR5984 at 5\,$\mu$m. The data were reduced applying procedures developed for the ESO Large Program 179.C-0151, as described in \citet{pont11}. The spectra were photometrically calibrated using the telluric standards as spectrophotometric references\footnote{Absolute fluxes for the standards were interpolated using photometry available through the VizieR catalogue, CDS, Strasbourg, France.}, achieving a precision of $\lesssim25$\% of the absolute fluxes (including flux uncertainties from the standard stars and variable atmospheric and slit transmission at the time of observations). We find that the continuum in EX Lupi decreased from $1.3\pm0.3$ Jy in April 2008 to $0.29\pm0.04$ Jy in April 2014 at 3\,$\mu$m, and from $4.2\pm0.6$ Jy to $0.34\pm0.03$ Jy at 5\,$\mu$m.

Together with a decrease in the continuum flux level, the NIR veiling decreased from 2008 to 2014. The CRIRES spectra of EX Lupi obtained in 2014 show a moderately veiled stellar photosphere that was not previously detectable in outburst (Figure \ref{fig:proposal}). Following \citet{muzer}, we estimate the veiling ($r_{\lambda} = EW_{ref}/EW - 1$, where $EW$ is the equivalent width of photospheric lines) by comparison to a reference spectrum of a main sequence standard star of similar spectral type. The standard star spectrum (HIP49986, M2V) was obtained using CRIRES and the same instrumental setup as for EX Lupi. We find a 3\,$\mu$m veiling of $r_{3}= 3.5 \pm 0.5$. At 5\,$\mu$m, no photospheric main sequence standards were observed, so we use TW Hya, a transitional disk \citep{calvet02} that is known to have low veiling \citep[$r_{5}=1.1$,][]{sal09}. By relative comparison, we find a veiling of $r_{5}= 4.0 \pm 1.5$ at 5\,$\mu$m for EX Lupi. We then determine the excess flux $F_{exc} = F_{\lambda} \times r_{\lambda} / (1+r_{\lambda})$, and estimate the temperature of the veiling dust by assuming blackbody emission. While an additional veiling measurement at 2\,$\mu$m is needed to improve the fit, we find the 3--5\,$\mu$m excess flux to be best represented by a temperature of $\sim$ 1200\,K, consistent with the lower end of values found in other T Tauri systems \citep[1200-1800\,K,][]{muzer,mcclure}.

\begin{figure}[ht]
\centering 
\includegraphics[width=0.44\textwidth]{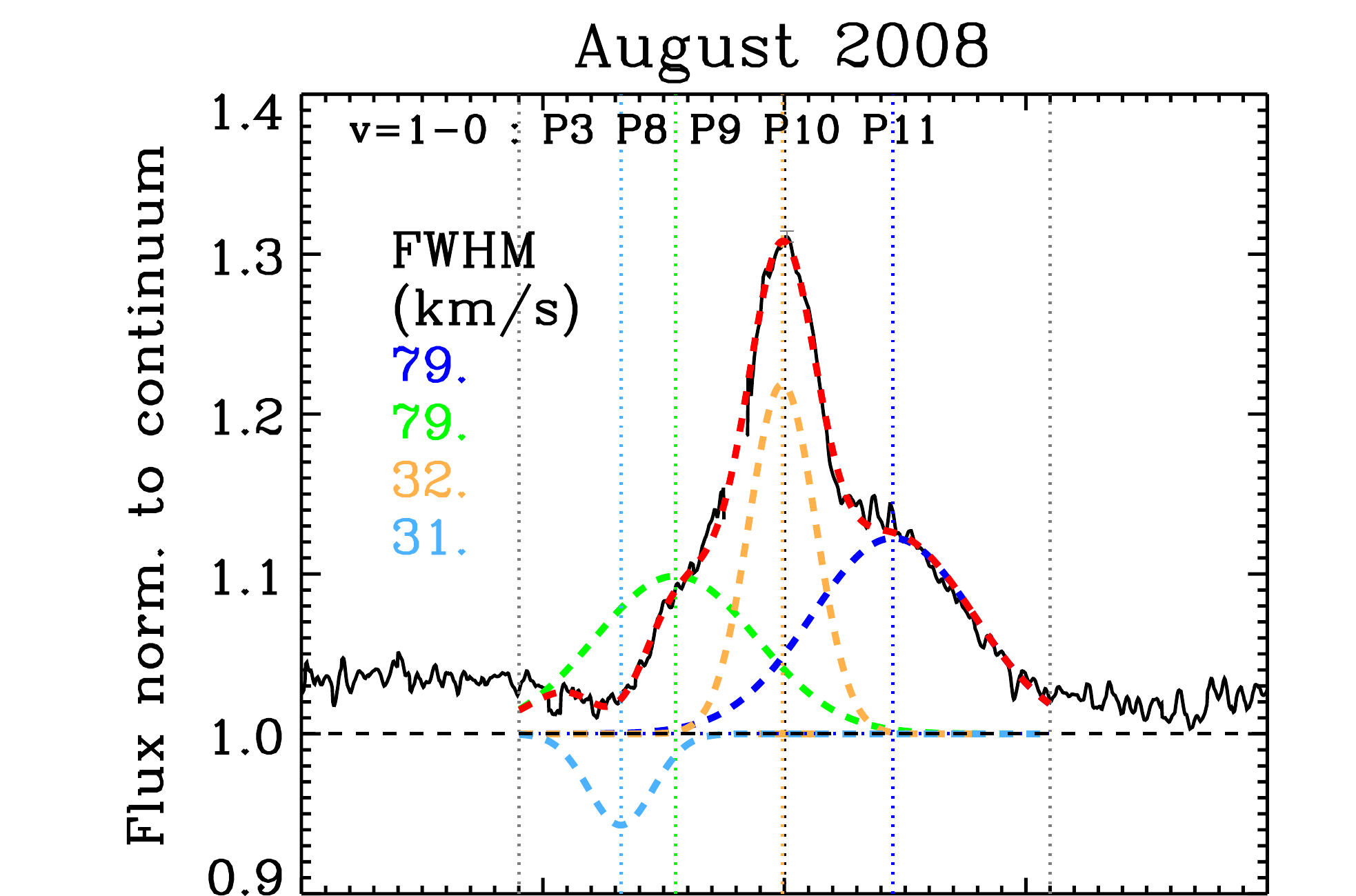} 
\includegraphics[width=0.44\textwidth]{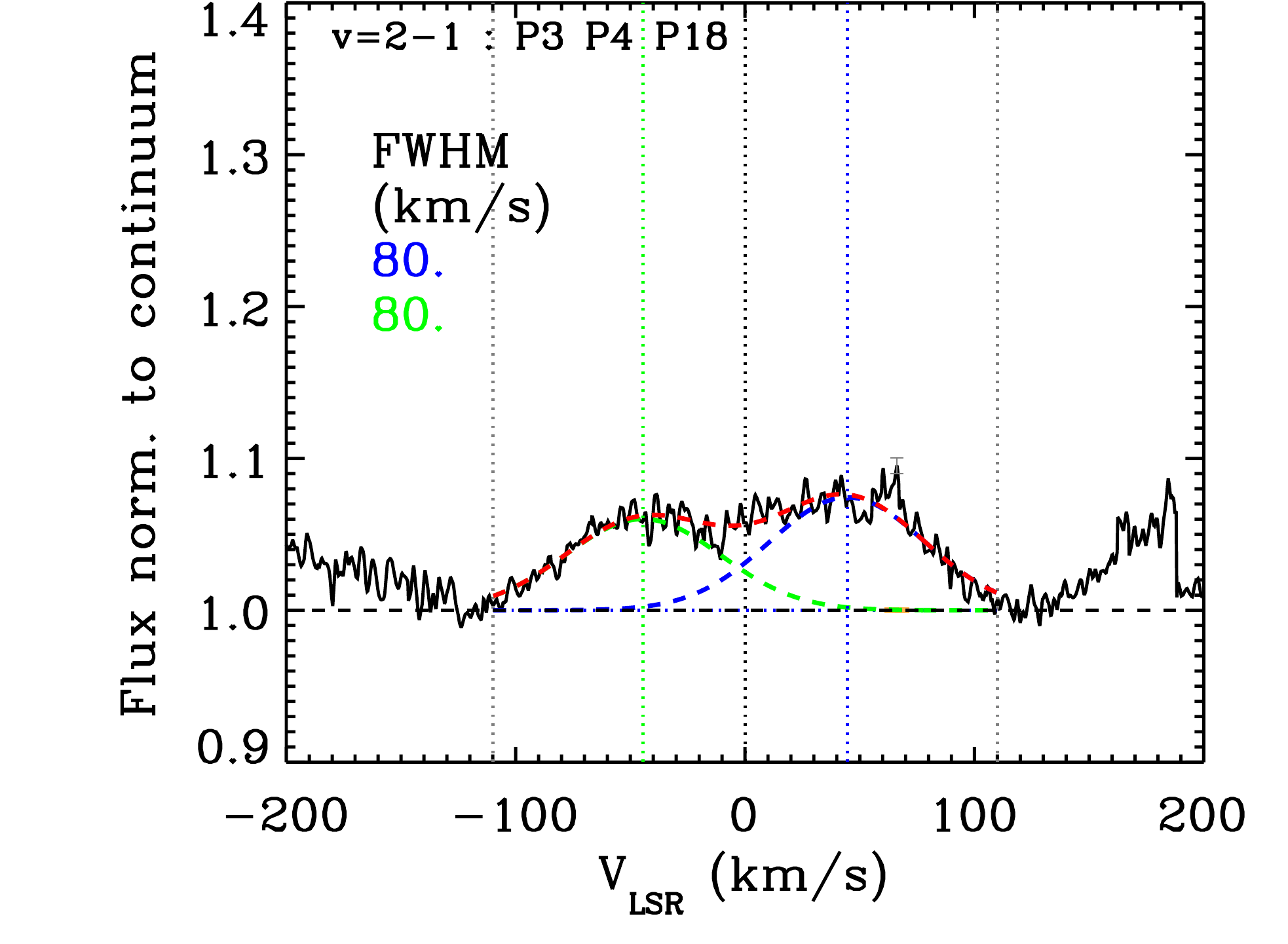} 
\includegraphics[width=0.44\textwidth]{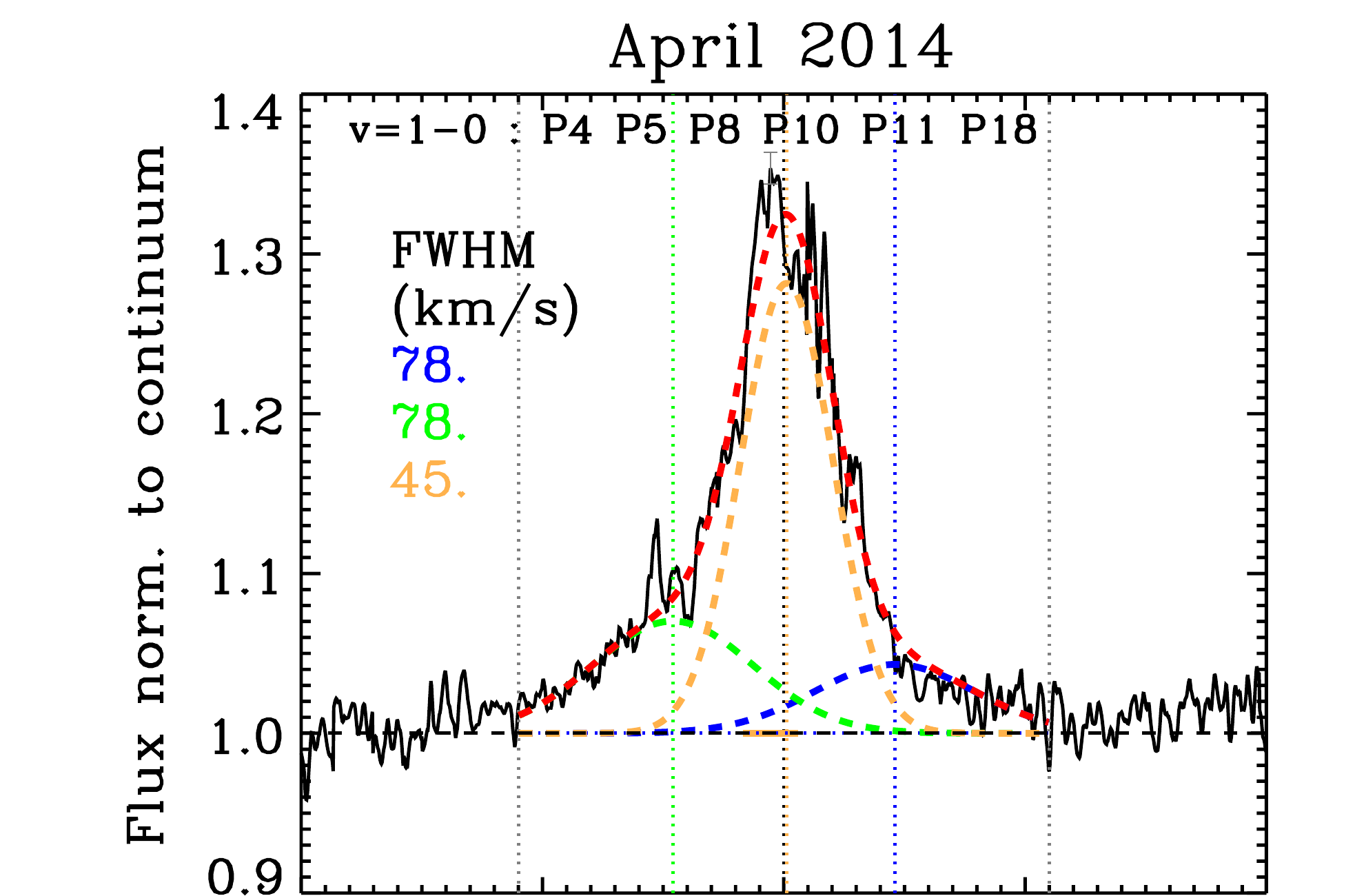} 
\includegraphics[width=0.44\textwidth]{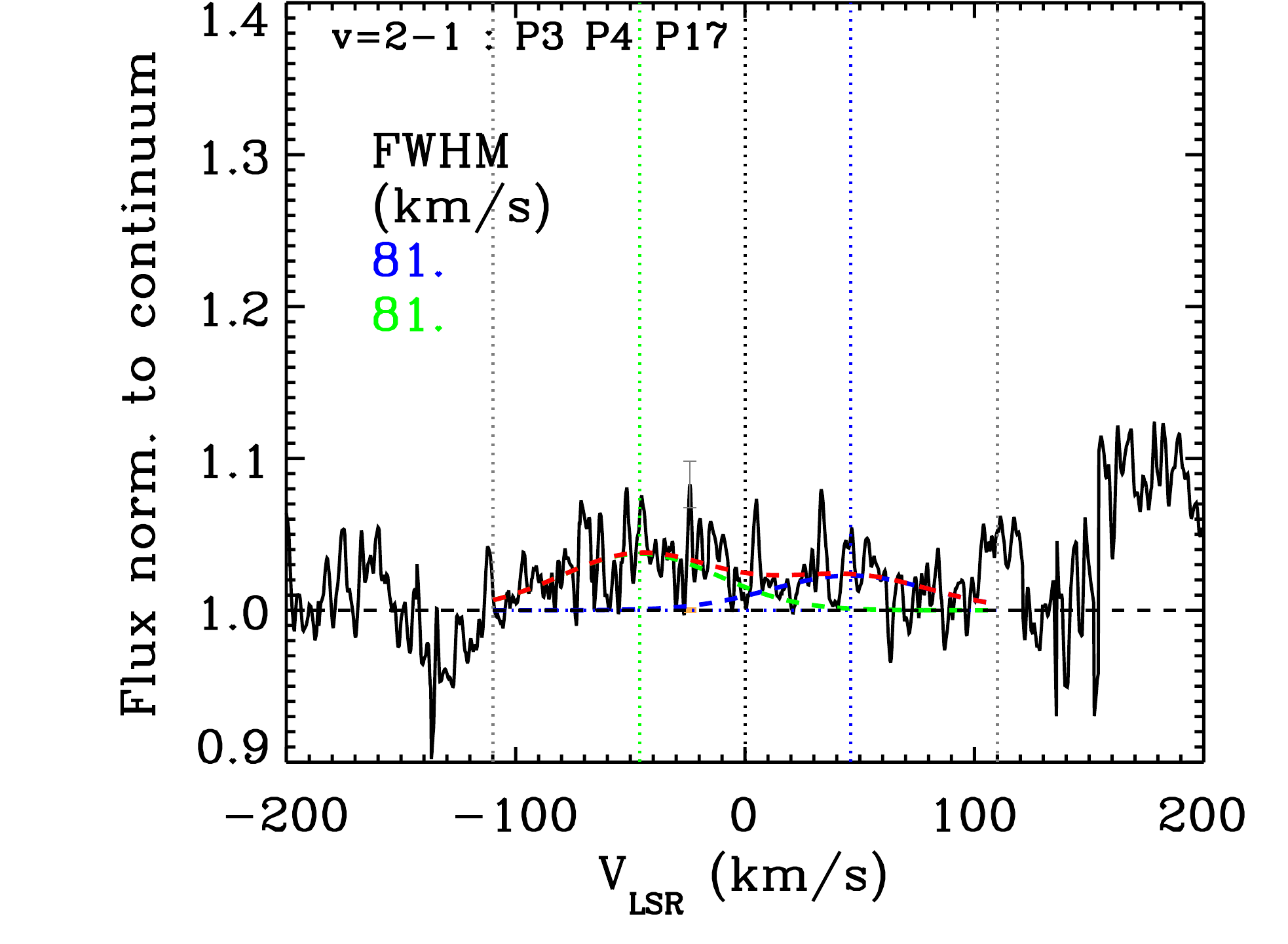} 
\caption{Averaged CO line profiles from the CRIRES spectra of EX Lupi. The individual lines used in the average are listed at the top of each plot. Gaussian fits to the different velocity components are shown as dashed colored lines overlaid to the data.}
\label{fig:line_width}
\end{figure}

\section{EVOLUTION OF THE MOLECULAR LINE EMISSION} \label{sec:resu}
 At NIR wavelengths, protoplanetary disks are known to show line emission from CO at 5\,$\mu$m \citep[e.g.][]{naji03}, and H$_2$O, OH, HCN, C$_2$H$_2$ at 3\,$\mu$m \citep[][]{sal08,mand12}. Since the CO spectra provide broad coverage of upper level energies and optical depths they provide the best constraints on the physical gas properties (Section \ref{sec:resu_1}). We apply the best-fit gas parameters from CO to the H$_{2}$O and OH spectra in Section \ref{sec:resu_2}.
 
\subsection{CO emission} \label{sec:resu_1}
Two velocity components of CO rovibrational lines are detected in emission at 5\,$\mu$m, both in 2008 and in 2014 (Figure \ref{fig:line_width}). A narrow component is detected only in $v=1-0$ lines, while a broad component of double-peaked lines is clearly detected in levels up to $v=4$ in 2008 \citep[see also][]{goto}. We characterize the line profile of the different velocity components by stacking CO lines between P3 and P18 to increase the signal-to-noise ratio. The number of lines we included in the average depends on the spectral coverage obtained, interrupted by detector chip gaps and telluric absorption, and on the level of contamination from higher vibrational levels. In 2008, neighboring lines from higher vibrational levels build up signal on top of the continuum at both sides of the stacked $v=1-0$ line profile, as seen in Figure \ref{fig:line_width}. We fit for the width, velocity separation, and peak flux ratio of two Gaussian functions in the broad CO $v=2-1$ lines. We then fix the best-fit velocity separation and peak ratio from the $v=2-1$ lines to the broad component of the $v=1-0$ lines, and fit for the widths and peaks of the broad and narrow components. An additional Gaussian needs to be included in absorption in 2008 to account for lack of emission at the blue side of $v=1-0$ lines; this blue-shifted absorption in EX Lupi had already been found by \citet{goto}, who attributed it to a disk wind developed during the outburst. 

We measure a full width at half maximum of $\sim$\,160 and 140\,km\,s$^{-1}$ for the broad component, and of $\sim$ 30 and 45\,km\,s$^{-1}$ for the narrow component in 2008 and 2014 respectively. The disk inclination in EX Lupi is still uncertain, but a value of $45^\circ$ was proposed by previous modeling of NIR CO emission \citep{goto,kosp11}. Assuming this inclination and gas in Keplerian rotation, and using the line velocity measured at 10\% of the line peak, the CO line widths translate into emitting inner disk radii $R_{in}$ of 0.02 AU for the broad component, and 0.2--0.3 AU for the narrow component.

\begin{figure*}[ht]
\centering
\includegraphics[width=0.49\textwidth]{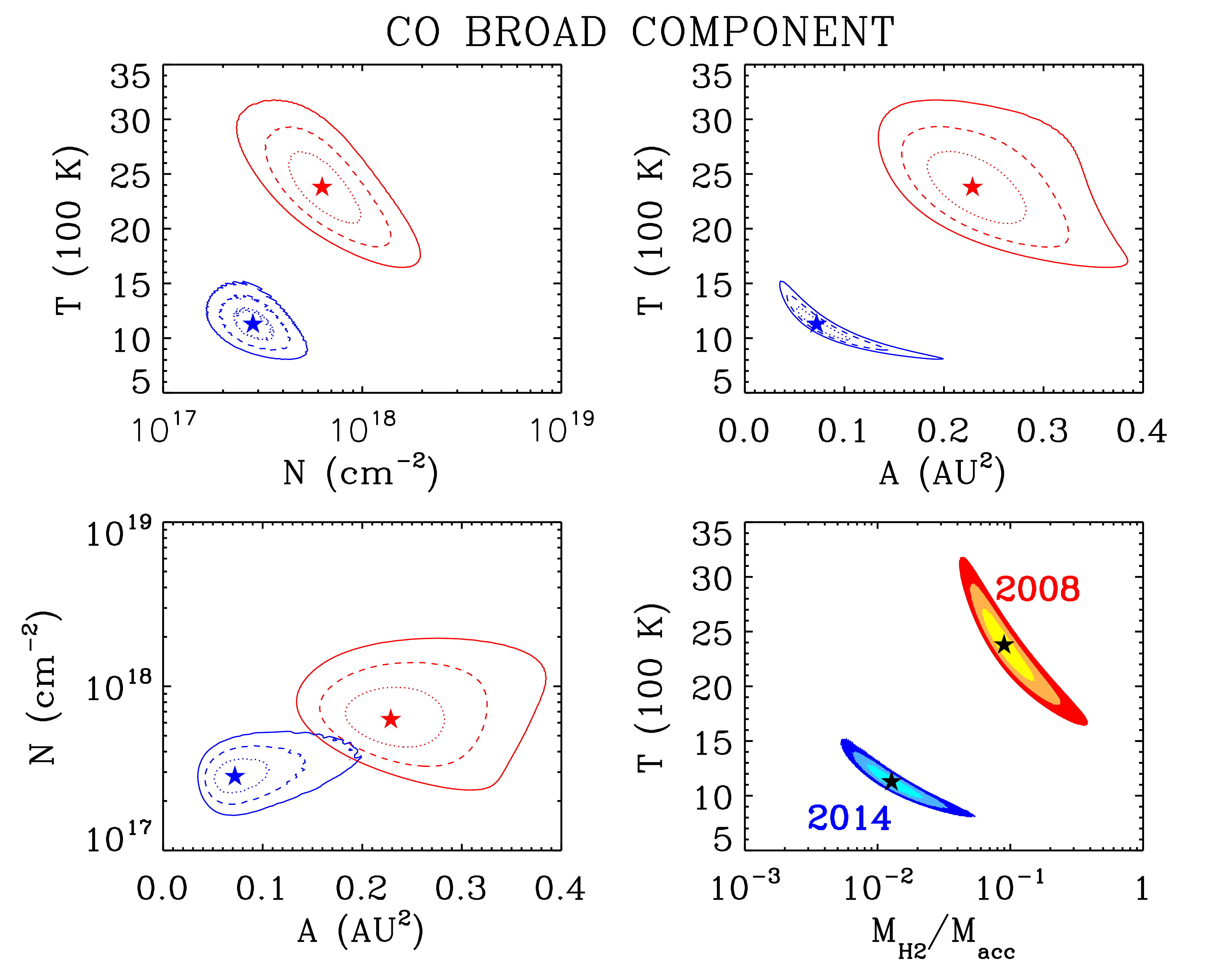} 
\includegraphics[width=0.49\textwidth]{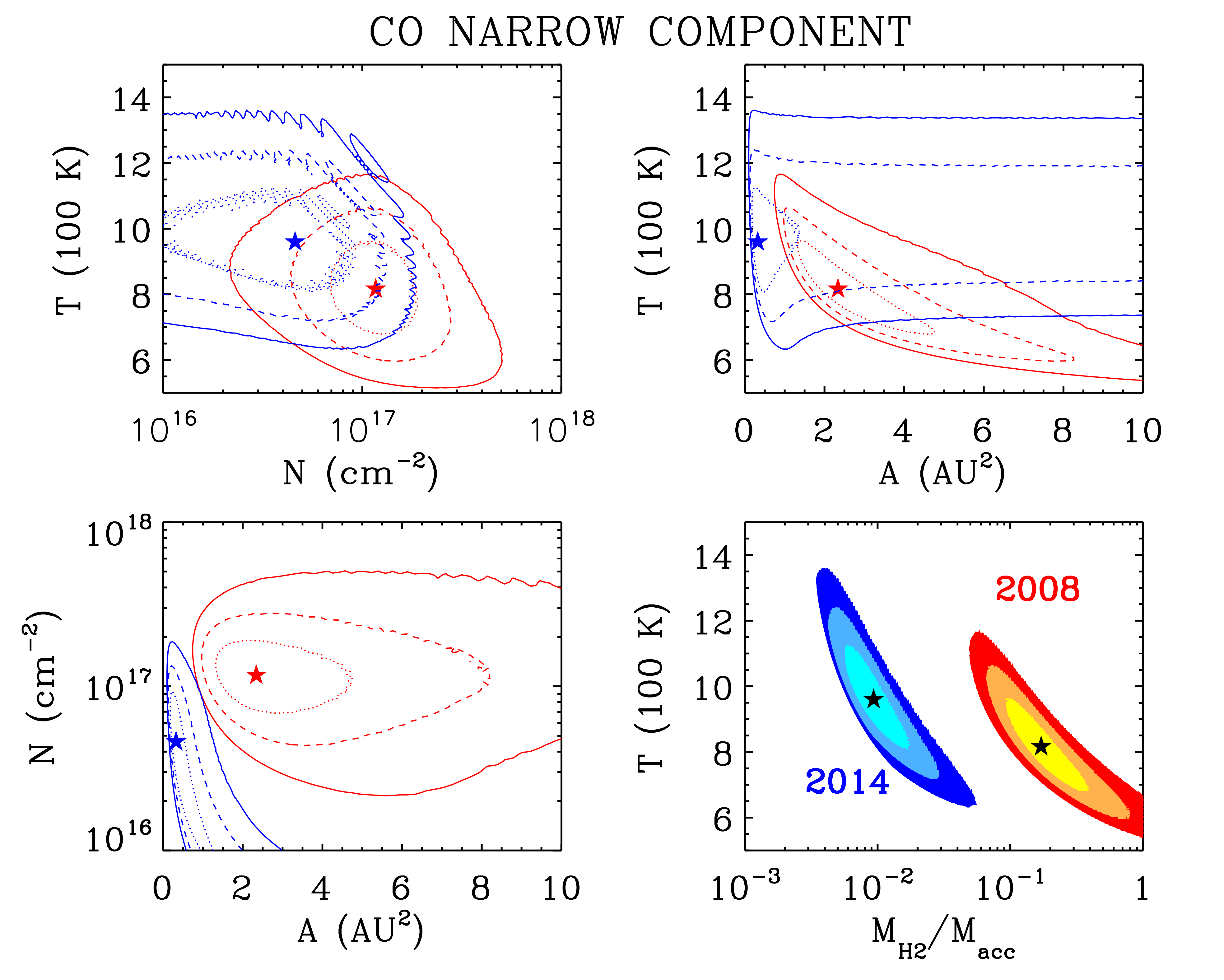} 
\includegraphics[width=0.49\textwidth]{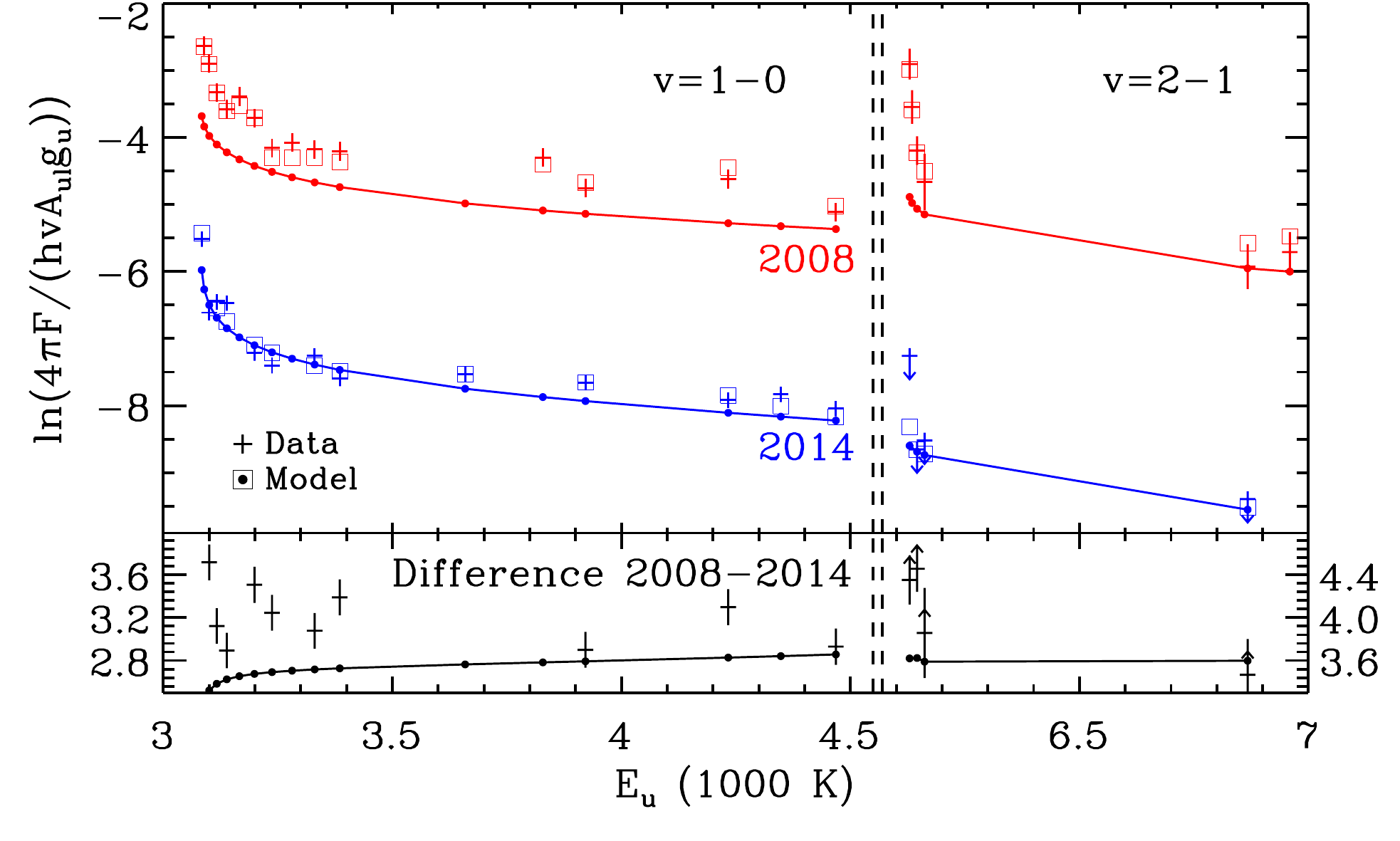} 
\includegraphics[width=0.49\textwidth]{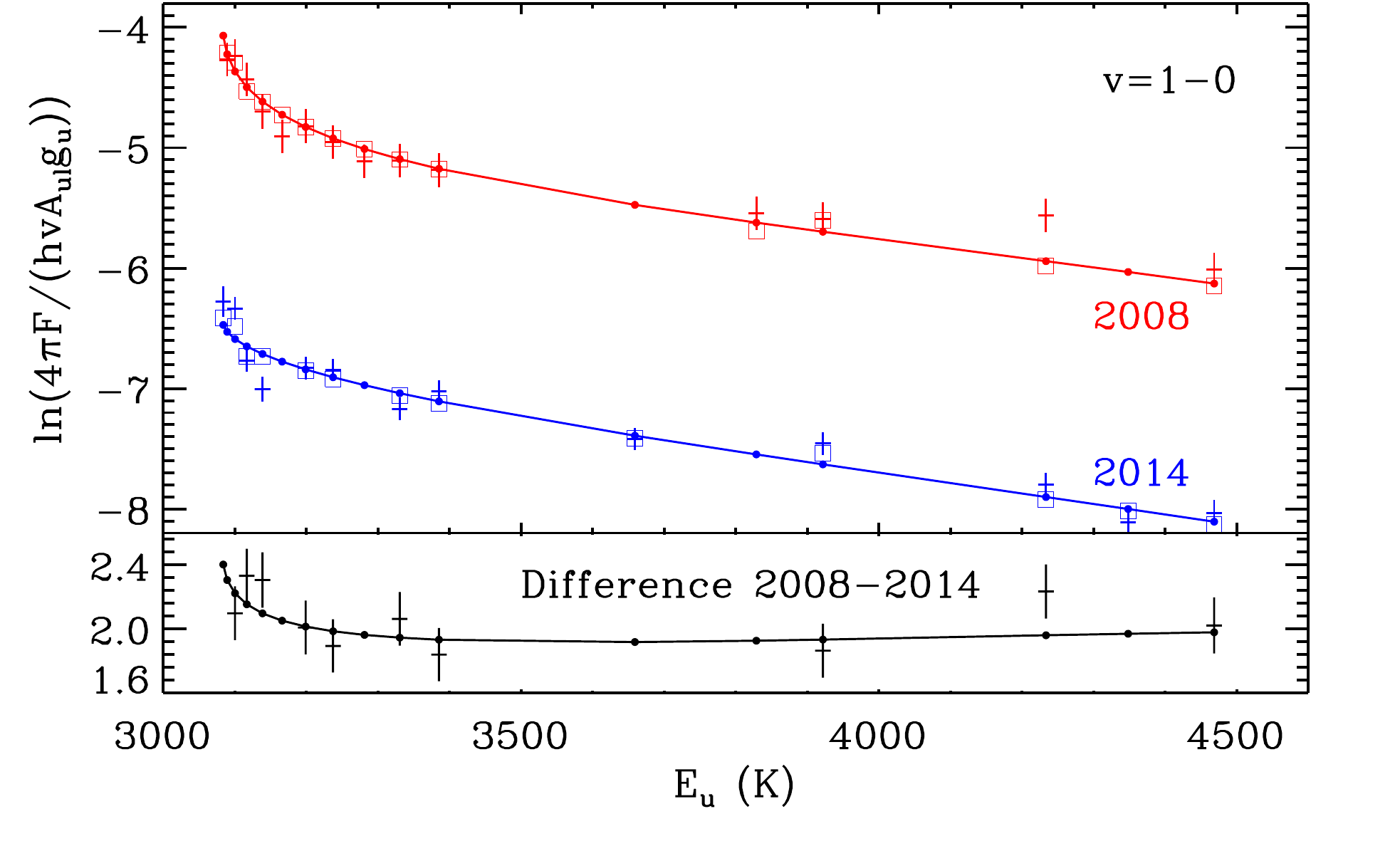} 
\caption{\textit{Top:} Two-dimensional confidence regions for the estimated model parameters (1, 2, and 3\,$\sigma$). Best-fit models are marked with a star. \textit{Bottom:} Population diagrams \citep{GL99} showing best-fit models overlaid to the data. Crosses show the measured line fluxes and their errors. Dots connected by a line show the unblended CO transitions from the models, while squares show the model fluxes blended with higher vibrational transitions as observed. Blending with $v\geq 2$ lines is most significant for the broad CO in 2008.}
\label{fig:chi2_lim}
\end{figure*}

To measure line fluxes of the two velocity components separately, we fit for the amplitudes of the different Gaussian functions in individual CO $v=1-0$ lines, after fixing the line widths and centroids from the fits to the stacked line profiles. To estimate the gas properties we then fit the extracted line fluxes with slab models accounting for line opacity, where the free parameters are the excitation temperature $T$, the column density $N$, and the emitting area $A$. The model uses the HITEMP and HITRAN molecular databases \citep[][]{hitemp,hitran12} and assumes thermal equilibrium, i.e. a Boltzmann distribution for the level populations \citep[an explicit description was included in][]{banz12}. We run grids of models to explore a large parameter space that includes the range of values found for NIR CO emission in previous studies. We include all measurable CO $v=1-0$, $2-1$, and $3-2$ line fluxes in the fits. We fix the vibrational temperature to the rotational temperature, and find reduced chisquare values $\leq1.6$ under this assumption. We find that both velocity components decrease in column density and emitting area from 2008 to 2014, by a factor of 2 and 3--8, respectively (Table\,\ref{tab:conf_lim} and Figure \ref{fig:chi2_lim}). In addition, the broad component decreases in temperature by a factor of 2 (as given by the decrease in $v=2-1$ to $v=1-0$ line ratios, Figure \ref{fig:chi2_lim}). The emission includes optically thin and thick lines in both epochs ($\tau=$\,0.05--6), so it can be used to estimate the total emitting CO gas mass as $N \times m_{\rm{co}} \times A$, where $m_{\rm{co}}$ is the mass of the CO molecule. Over the same area, the H$_{2}$ gas mass $M_{\rm{H}_{2}}$ is estimated by assuming a fractional abundance of CO/H$_{2} = 10^{-4}$, as found appropriate for the inner regions of a disk \citep[][]{france}. In Figure\,\ref{fig:chi2_lim} we show $M_{\rm{H}_{2}}$ in units of the total mass accreted during outburst $M_{\rm{acc}}=6 \times 10^{-9}$\,M$_{\odot}$, estimated using an average accretion rate of $10^{-8}$\,M$_{\odot}$/yr \citep{sicagu12} over seven months of the outburst. We find that $M_{\rm{H}_{2}} / M_{\rm{acc}} \sim$ 0.1--1 in outburst\footnote{The 2008 CO spectrum was taken late during the outburst, when some fraction of the gas mass had already been accreted.}, and that $M_{\rm{H}_{2}}$ decreased by roughly one order of magnitude from August 2008 to 2014 in both the broad and narrow components.

\begin{deluxetable}{l l l l}
\tablecaption{ \label{tab:conf_lim}}
\tablewidth{0pt}
\tabletypesize{\small}
\tablehead{\colhead{Line Sample} & \colhead{Parameter} & \colhead{2008} & \colhead{2014}}
\startdata
  \vspace{5pt}
& \textit{$R_{\rm in}$} (AU) & $0.02$ &  $0.02$ \\ %
  \vspace{5pt}
& \textit{$T_{\rm ex}$} (K) & $2400^{+200}_{-200}$ &  $1100^{+100}_{-100}$ \\ %
 \vspace{5pt}
CO broad & \textit{$N_{\rm mol}$} ($10^{17}$\,cm$^{-2}$) & $6.3^{+2.0}_{-1.4}$ &  $2.8^{+0.4}_{-0.4}$ \\ %
 \vspace{5pt}
 & \textit{$A$} (AU$^{2}$) & $0.23^{+0.03}_{-0.03}$ & $0.07^{+0.02}_{-0.01}$  \\
 & \textit{$M_{\rm H_{2}}$} ($10^{-10}$\,M$_{\odot}$) & $5.4^{+1.9}_{-1.2}$ & $0.8^{+0.2}_{-0.2}$  \\ 
\\

\hline
\\
  
  \vspace{5pt}
& \textit{$R_{\rm in}$} (AU) & $0.30$ &  $0.17$ \\ %
  \vspace{5pt}
& \textit{$T_{\rm ex}$} (K) & $800^{+100}_{-100}$ & $960^{+100}_{-100}$ \\
 \vspace{5pt}
CO narrow & \textit{$N_{\rm mol}$} ($10^{17}$\,cm$^{-2}$) & $1.2^{+0.4}_{-0.3}$ & $0.5^{+0.3}_{-0.3}$ \\
 \vspace{5pt}
 & \textit{$A$} (AU$^{2}$) & $2.3^{+1.3}_{-0.7}$ & $0.3^{+0.3}_{-0.1}$  \\
 & \textit{$M_{\rm H_{2}}$} ($10^{-10}$\,M$_{\odot}$) & $10.3^{+4.9}_{-2.7}$ & $0.6^{+0.1}_{-0.1}$  
  
\enddata
\tablecomments{$R_{\rm in}$ is derived from line widths, using a disk inclination of $45^\circ$ (Section \ref{sec:resu_1}); other parameters come from slab model fits. Errors show the 1\,$\sigma$ confidence on each parameter estimate.} 
\end{deluxetable}

\subsection{H$_2$O and OH emission} \label{sec:resu_2}
Emission lines from H$_{2}$O and OH are detected only in the outburst spectrum, at 3\,$\mu$m (Figure \ref{fig:proposal}). They have line widths comparable to the broad component of the $\Delta v=1$ CO lines, suggesting that they trace molecular gas extending inward to similar disk radii (0.02\,AU). The H$_{2}$O spectrum is comparable to that seen in a few other T Tauri systems so far \citep{sal08,mand12}, although higher vibrational states are populated and the lines are significantly broader, therefore suffering a greater degree of blending. We apply the best-fit temperature and area found for the broad CO component to H$_{2}$O and OH emission, and to model the 3\,$\mu$m spectrum we need column density ratios of CO/H$_{2}$O\,$\sim10$ and OH/H$_{2}$O\,$\sim1$, in the range found by \citet{mand12} in other disks. The broad CO component and the H$_{2}$O and OH gas therefore may have undergone the same depletion and cooling from 2008 to 2014, explaining the disappearance of line emission at 3\,$\mu$m.

\section{Discussion} \label{sec:disc}
From the evolution in NIR CO emission observed by comparison of CRIRES spectra of EX Lupi, we estimated that $M_{\rm{H}_{2}}$ in the inner disk decreased by one order of magnitude from August 2008 to April 2014. EX Lupi underwent a major accretion event in 2008, when the accretion rate onto the star increased by two orders of magnitude to $10^{-8}$--$10^{-7}$\,M$_{\odot}$\,yr$^{-1}$ \citep[Figure \ref{fig:proposal}, and][]{asp}. However, the system spends most of its time in a quiescent state with a much lower accretion rate \citep{herb01,niki}. \citet{sicagu12} estimated from monitoring of the H$\alpha$ 10\% width that before and after the 2008 outburst EX Lupi had a similar low accretion rate of $\approx10^{-10}$\,M$_{\odot}$\,yr$^{-1}$. If the 2014 data show the typical conditions of the quiescent system, what we measure is a higher gas mass in the inner disk during outburst (2008) by comparison to quiescence. 

This behavior is consistent with the current view of episodic accretion in young stars \citep[see e.g. review by][]{audard14}. In particular, one set of models have been shown to be relevant for the timescales and strengths of EXor outbursts \citep[][]{spruit,dAng10,dAng11,dAng12}. In these models, when the inner disk of a T Tauri star is truncated by the stellar magnetosphere outside (but close to) the corotation radius $R_{c}$, the accretion rate onto the star is inhibited and gas accumulates beyond $R_{c}$. When the surface density becomes large enough to overcome the centrifugal barrier, the magnetospheric radius $R_{m}$ is perturbed inward of $R_{c}$ and accretion onto the star can proceed vigorously until the gas reservoir is emptied, eventually causing $R_{m}$ to migrate outward of $R_{c}$ again into a low-accretion phase. \citet{dAng12} showed that this accumulation-release behavior can lead to excursions in the accretion rate of about two orders of magnitude and outburst duration of months, consistent with observations of EX Lupi (see cartoon in Figure \ref{fig:proposal}). Using the definition in \citet{bouvier}, the radius $R_{m}$ where the dipolar field in the stellar magnetosphere truncates the disk in EX Lupi should be located at 0.2-0.3 AU in 2014 and at 0.03-0.06 in 2008 \citep[adopting a magnetic field strength between 1 and 3\,kG as appropriate for T Tauri stars,][]{JK07}. The corotation radius, where the disk rotation equals the stellar rotation, is instead at $R_{c}\sim0.05$\,AU or outward at $R_{c}\gtrsim0.07$\,AU \citep[estimated using a stellar $v \sin i=4.4$ and $<3$\,km\,s$^{-1}$ from][respectively]{niki,kosp14}. Although all these estimates are uncertain, the conditions for the \citet{dAng12} model may therefore be found in EX Lupi, as $R_{m}$ is larger than $R_{c}$ in 2014 (when accretion onto the star is inhibited) and smaller than $R_{c}$ during outburst (when the accretion instead proceeds vigorously). The broad CO component traces the region at and inward of $R_{c}$, as found by \citet{sal11b} for CO emission in other T Tauri disks; in the \citet{dAng12} scenario, the increase in gas mass by an order of magnitude would measure the release of gas when $R_{m}$ approaches $R_{c}$. The narrow CO component, instead, may probe the gas accumulation/feeding region beyond $R_{c}$. Indeed, during the outburst $M_{\rm{H}_{2}} / M_{\rm{acc}}$ approaches unity, while in 2014 $M_{\rm{H}_{2}}$ is just sufficient to feed a quiescent accretion rate of $\sim10^{-10}$\,M$_{\odot}$\,yr$^{-1}$. $R_{m}$ in 2014 is very similar to $R_{in}$ of the narrow CO lines, and may explain the reason of two physically separated emission components as due to the inner disk truncation by the stellar magnetosphere. An $R_{in}<R_{m}$ for the broad component in 2014 may present an inconsistency with this picture, unless $R_{m}$ can be put to 0.02\,AU by the combination of dipolar and octupolar field components \citep{gregory,adams}. To test the validity of this interpretation of the CO emission seen in EX Lupi, future monitoring should measure a steady gas mass from the CO broad component (until the next outburst occurs) and an increasing gas mass in the narrow component as gas accumulates during quiescence. Monitoring variations in the NIR CO emission observed from EXors could provide unique tests of episodic accretion models.

\begin{figure}
\centering
\includegraphics[width=0.5\textwidth]{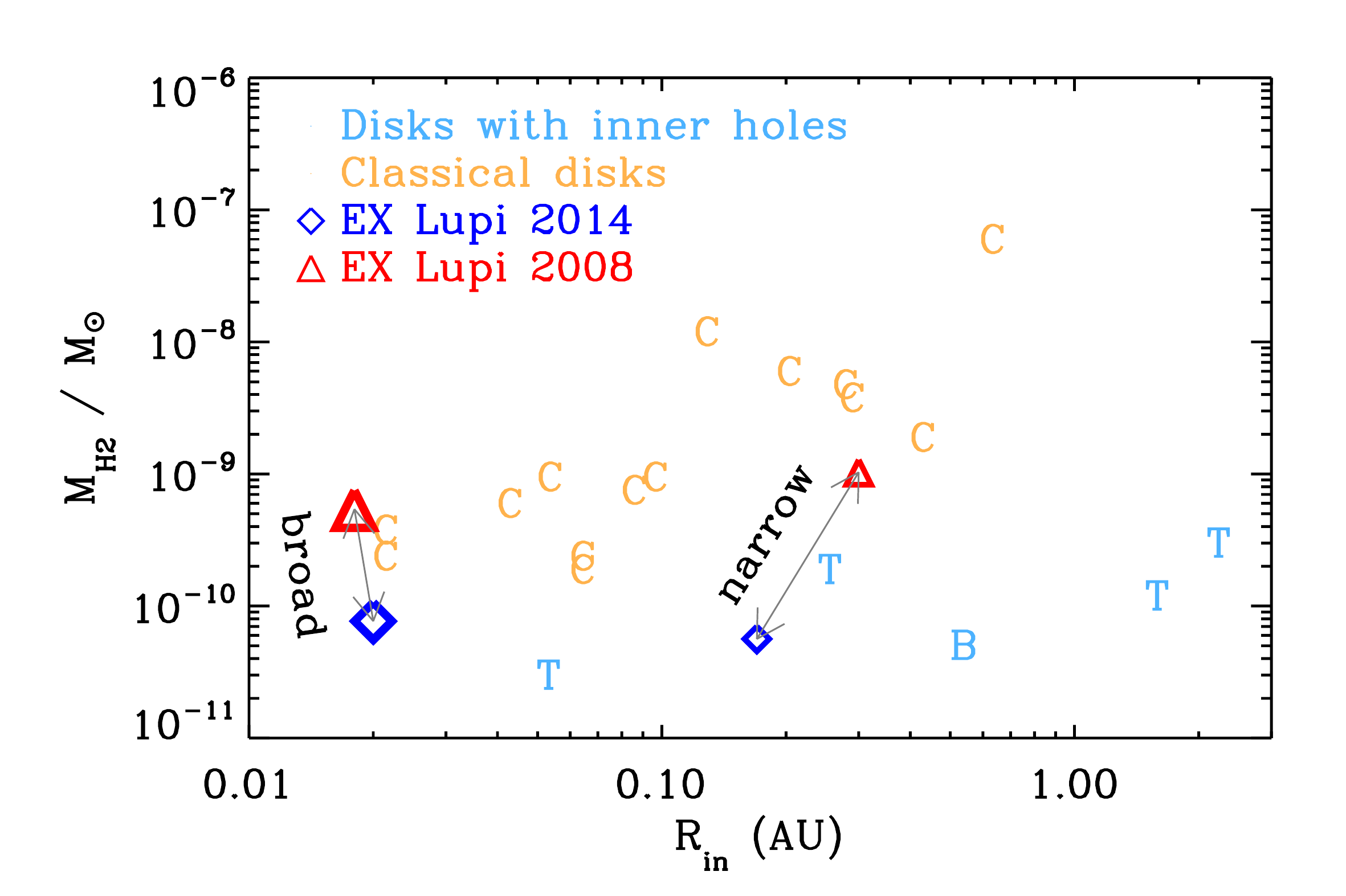} 
\caption{Changes in gas emission in EX Lupi between 2008 and 2014 (broad/narrow component in large/small symbols). Values for ``classical" (C) and ``transitional'' (T) are taken from \citet{sal11b}. $M_{\rm{H}_{2}}$ is estimated from $N$ and $A$ as in Section \ref{sec:resu_1}. The circumbinary disk (B) is taken from \citet{carr01}.}
\label{fig:discussion}
\end{figure}

EX Lupi should also be discussed in the context of similarly young (1-5 Myr) disks. In Figure \ref{fig:discussion} we compare the gas properties as measured from NIR CO emission in EX Lupi to individual values found in other young disks by \citet{sal11b}, who used methods comparable to those used in this work. We include in the figure both CO velocity components from EX Lupi, although a multi-component analysis has not been performed yet in other disks. The narrow CO emission in EX Lupi shows properties well within those found in other disks, while the broad emission lies at the lower extreme in $R_{in}$ values. In 2008, the gas in EX Lupi reached masses consistent with those of classical T Tauri disks. In 2014, instead, EX Lupi lies within the systematically lower values found in ``transitional" and circumbinary disks, i.e. disks that show dissipation of gas and dust outward of the dust sublimation radius. An inner dust radius larger than sublimation has been proposed for EX Lupi by \citet{niki} and \citet{attila}, at a location consistent with $R_{in}$ of the narrow CO lines (0.2--0.3\,AU), and a close-in (0.06\,AU) low-mass ($m\sin i\sim15$\,M$_{\rm{Jup}}$) binary companion has been recently claimed by \citet{kosp14}. Conclusive evidence is still awaited, and our findings raise questions on EX Lupi that may apply to episodically accreting T Tauri systems in general. Does a link exist between accretion outbursts and the dissipation status of inner disks? Is EX Lupi a classical T Tauri system that resembles a transitional disk in its inner disk gas properties, or a transitional disk that looks classical only during accretion outbursts? As a step toward answering such questions, we have found that high-resolution infrared spectroscopy of molecular gas helps to determine the dissipation and evolution of inner disks, even in regimes where these are insufficient to produce significant signatures in the broad-band spectral energy distribution.

The authors thank referee Greg Herczeg for comments and suggestions that helped improving this work. A.B. acknowledges financial support by a NASA Origins of the Solar System Grant No. OSS 11-OSS11-0120, a NASA Planetary Geology and Geophysics Program under grant NAG 5-10201. M.R.M. acknowledges financial support of the Swiss National Science Foundation within the framework of the National Centre for Competence in Research ``PlanetS". This work is based on observations made with ESO telescopes at the Paranal Observatory under programs 093.C-0432 and 179.C-0151.

\end{document}